# PET CMR$_{glc}$ mapping and $^1$H-MRS show altered glucose uptake and neurometabolic profiles in BDL rats

Jessie Mosso[1,2,3], Ting Yin[1,2], Carole Poitry-Yamate[1], Dunja Simicic[1,2,3], Mario Lepore[1,2], Valérie A. McLin[4], Olivier Braissant[5], Cristina Cudalbu[1,2], and Bernard Lanz[3]

[1]CIBM Center for Biomedical Imaging, Switzerland

[2]Animal Imaging and Technology (AIT), EPFL, Lausanne, Switzerland

[3]Laboratory for functional and metabolic imaging (LIFMET), EPFL, Lausanne, Switzerland

[4]Swiss Pediatric Liver Center, Department of Pediatrics, Gynecology and Obstetrics, University Hospitals Geneva, and University of Geneva, Geneva, Switzerland

[5]Service of Clinical Chemistry, Lausanne University Hospital and University of Lausanne, Lausanne, Switzerland

**Corresponding author**: bernard.lanz@epfl.ch

**Abbreviations**:

ADP : Adenosine diphosphate

AIF : arterial input function

Ala : Alanine

Asc : ascorbate

Asp : aspartate

ATP : Adenosine triphosphate

BBB : blood brain barrier

BDL : bile duct ligated

bHB : β-hydroxybutyrate

$CMR_{glc}$ : cerebral metabolic rate of glucose

Cr : creatine

FDG : fluorodeoxyglucose

FDG6P : fluorodeoxyglucose-6-phosphate

FOV : field of view

G6P : glucose-6-phosphate

GABA : γ-aminobutyric acid

Glc : glucose

Gln : glutamine

Glu : Glutamate

Glu : glutamate

GPC : glycerophosphocholine

GS : glutamine synthetase enzyme

GSH : glutathione

HE: hepatic encephalopathy

Ins : myo-inositol

Lac : lactate

LC : Lumped Constant

MIP : maximum intensity projection

MLEM : maximum likelihood expectation maximization

NAA : N-acetylaspartate

NAAG : Nacetylaspartylglutamate

PCho : phosphocholine

PCr : phosphocreatine

PE : phosphoethanolamine

PET : positron emission tomography

ROI : region of interest

Scyllo : scyllo-inositol

SD : standard deviations

SUV : standardized uptake value

SyN : symmetric deformable registration

Tau : taurine

Tau : taurine

tCho : total choline

tCr : total Creatine

$TE_{eff}$ : effective echo time

tNAA : total *N*-acetylaspartate

TR : repetition time

VOI : volumes of interest

$\gamma$-ATP : gamma adenosine triphosphate

# Graphical abstract

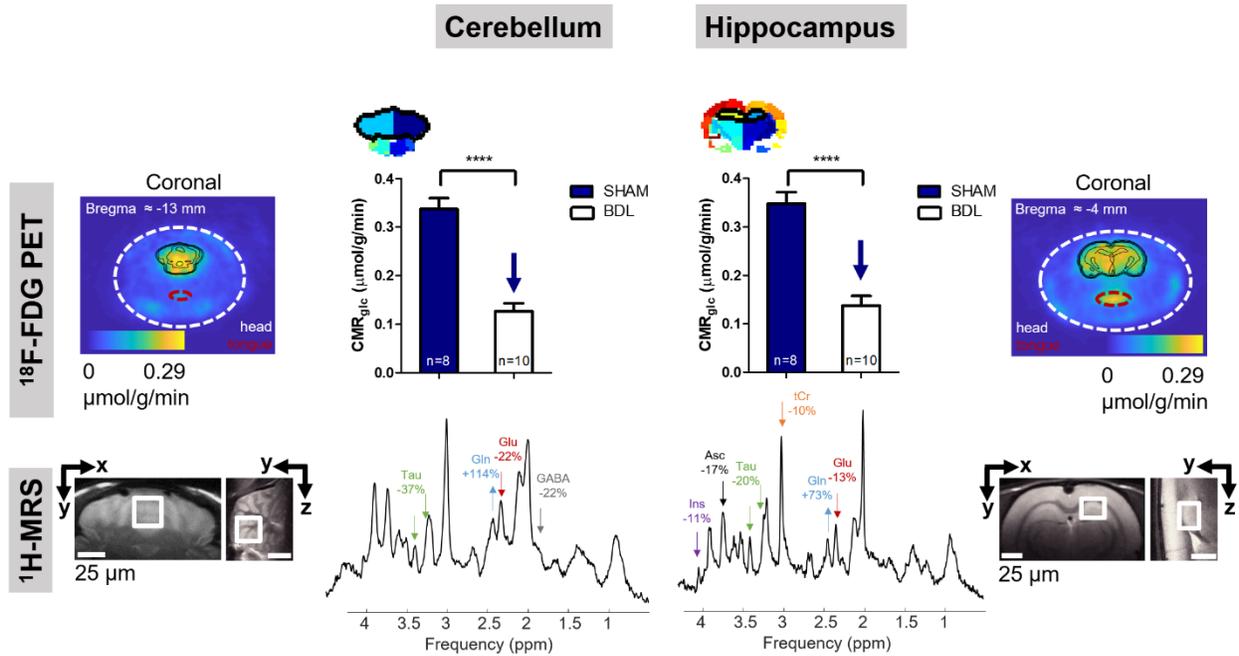

# Abstract

Type C hepatic encephalopathy (HE) is a complex neuropsychiatric disorder occurring as a consequence of chronic liver disease. Alterations in energy metabolism have been suggested in type C HE, but *in vivo* studies on this matter remain sparse and have reported conflicting results. Here, we propose a novel preclinical $^{18}$F-FDG PET methodology to compute quantitative 3D maps of the regional cerebral metabolic rate of glucose (CMR$_{glc}$) from a labelling steady-state PET image of the brain and an image-derived input function. This quantitative approach shows its strength when comparing groups of animals with divergent physiology, such as HE animals. PET CMR$_{glc}$ maps were registered to an atlas and the mean $CMR_{glc}$ from the hippocampus and the cerebellum were associated to the corresponding localized $^1$H-MR spectroscopy acquisitions. This study provides for the first time local and quantitative information on both brain glucose uptake and neurometabolic profile alterations in a rat model of type C HE. A 2-fold lower brain glucose uptake, concomitant with an increase in brain glutamine and a decrease in the main osmolytes was observed in the hippocampus and in the cerebellum. These novel findings are an important step towards new insights into energy metabolism in the pathophysiology of HE.

## Introduction

Type C encephalopathy (type C HE) is a severe neuropsychiatric disorder occurring as a consequence of chronic liver disease, for which the prognosis is poor in the absence of liver transplantation[1]. In Type C HE, cirrhosis with systemic shunting not only blocks blood flow; it hinders the liver's ability to filter and detoxify natural toxins, like ammonium, out of the body, thus facilitating toxin build-up in systemic blood travelling to the brain, where it adversely affects brain function. The understanding of biochemical mechanisms underpinning neurological and cognitive dysfunctions is still incomplete. So far, ammonium ($NH_4^+$) accumulation and glutamine (Gln) metabolism have been considered to play a central role in the pathophysiology of type C HE. Due to improper toxin clearance by the diseased liver, $NH_4^+$ accumulates in the systemic circulation before reaching the brain. Elevation in brain ammonia concentrations leads to an excessive synthesis of brain Gln by the glutamine synthetase enzyme (GS), predominantly located in the astrocytes[2]. An increased astrocytic Gln content triggers an osmotic regulation mechanism shown by decreased concentrations of myo-inositol (Ins), taurine (Tau), total choline (tCho) and total Creatine (tCr) measured by *in vivo* $^1$H spectroscopy, and consequently low-grade brain edema[3–5]. Alterations in energy metabolism have also been investigated *in vivo* in type C HE, yet only small or undetectable changes have been reported[6]. In the adult bile duct ligated rat model of type C HE[7], no changes in reliably-quantified brain metabolites involved in energy metabolism (i.e. lactate (Lac), gamma adenosine triphosphate (γ-ATP)) have been observed using *in vivo* proton or phosphorus magnetic resonance spectroscopy[5,8,9] ($^1$H or $^{31}$P MRS). Using carbon ($^{13}$C) MRS after $^{13}$C-labeled Glc injection, a steady-state increased Lac pool has been detected in bile duct ligated (BDL) rats with uniformly labelled glucose (Glc)[10], but no changes in brain mitochondrial fluxes have been measured in the same animal model following [1,6-$^{13}$C$_2$] Glc injections[11]. In the wider scope of energy metabolism studies in other preclinical models of HE (i.e. chronic portacaval-shunted rats and urease-induced hyperammonemic rats), autoradiography studies have reported conflicting results[12,13]. In addition, studies in patients with HE using positron emission tomography (PET) studies have focused on cirrhotic patients with mild HE only, and have shown a decreased $^{18}$F fluorodeoxyglucose ($^{18}$F-FDG) uptake in the cingulate gyrus[14,15] and a hypermetabolism in the hippocampus[16,17]. A clinical case study of a patient with decompensated liver cirrhosis also showed a hypometabolism of glucose in the cerebellum and cerebral cortices[18]. However, to the best of our knowledge, no $^{18}$F-FDG PET studies on preclinical models of type C HE are available to date.

*In vivo* localized $^1$H MRS allows the non-invasive measure of metabolites involved in a variety of brain functions, such as osmoregulation (Ins, tCr), neurotransmission (Gln, Glutamate (Glu), γ-aminobutyric acid (GABA)) or energy metabolism (Lac). In the context of HE, *in vivo* longitudinal localized $^1$H MRS has been acknowledged as a predictive tool of the early stages of the disease [19,20]. Moreover, when *in vivo* localized $^1$H MRS is performed at ultra-high field (≥ 7 Tesla), the separation between Gln and Glu spectral peaks is feasible, allowing to disentangle the behaviour of these two crucial metabolites in the development of HE[5,9].

While $^1$H MRS provides a steady-state information on metabolic pools, $^{18}$F-FDG PET provides kinetic information on local brain glucose uptake. $^{18}$F-fluorodeoxyglucose (FDG), an analog of glucose labelled with the positron-emitting $^{18}$F, is transported across the blood-brain barrier and converted to $^{18}$F-FDG-6-phosphate (analog of glucose-6-phosphate (G6P)) and no further metabolized through the glycolysis[21]. Therefore, the FDG PET signal at labelling steady-state reflects the very first two steps of glycolysis: glucose transport through the blood brain barrier (BBB) and phosphorylation into G6P. The standardized uptake value (SUV) is routinely reported in PET studies, but it does not allow for a quantitative assessment of glucose cerebral metabolic rates, especially in pathological conditions with alterations of systemic metabolism. The derivation of the glucose cerebral metabolic rate ($CMR_{glc}$) from labelling steady-state images is, on the contrary, a quantitative method introduced by *Sokoloff et al.*[21] for 2-deoxy-D-[14C] glucose autoradiography studies, which can be extended to *in vivo* [18]F-FDG PET data and provides both a local and quantitative rate of glucose utilization in the tissue in a non-invasive way. Yet, it requires the dynamic measurement of the blood FDG activity (the arterial input function (AIF)) from the time of the bolus injection up to the labelling steady-state measurement time frame [22]. The measurement of arterial input function can be particularly challenging in rodents and is often the more invasive part of the FDG-PET study. Following a recent strategy proposed by *Lanz et al.*[23], the AIF can be measured prior to the brain acquisition from the PET image of the vena cava where the FDG bolus is observed. In doing so, difficulties linked to manual and multiple blood samplings can be circumvented. By combining this dynamic measurement of the AIF with a static PET measurement on the brain at labelling steady-state, the $CMR_{glc}$ can be derived from the sole PET scan for each animal with minimal invasiveness.

In the present work, we study how HE affects the first metabolic step in brain energetics, i.e. glycolysis, the uptake and metabolism of blood-derived glucose. To this end, and presented in full, a novel [18]FDG PET-based methodology was developed to quantitate regional $CMR_{glc}$, using the image-derived AIF. The proposed [18]F-FDG PET quantification method results in 3D spatial mapping of glucose uptake in micromole/g/min. In addition, we combined the $CMR_{glc}$ maps with *in vivo* [1]H MRS at 9.4T in the hippocampus and cerebellum, using a PET-to-atlas registration via intermediate MRI anatomical image co-registration, and showed regional alterations of brain glucose uptake in the BDL rat model of type C HE concomitant with neurometabolic pools' changes. We focused on the hippocampus and cerebellum, as they are key regions involved in HE[24,25]. Taken together, [1]H MRS and quantitative [18]F-FDG PET bring a new insight on brain energy metabolism in HE.

## Methods

**BDL rats:** We used the BDL rat model for chronic liver disease leading to type C HE, recognized by the International Society for Hepatic Encephalopathy and Nitrogen

Metabolism (ISHEN)[7]. Male adult Wistar rats (n=18, Charles River Laboratories, France) underwent BDL (n=10, 175 ± 13 g at surgery) or SHAM surgery (n=8, 174 ± 14 g at surgery). Plasma bilirubin (Reflotron Plus analyzer, Roche, Switzerland) and blood ammonia (Integra 400 Plus, Roche, Switzerland) from the sublingual vein were measured at week 6 post-surgery in BDL rats to follow the disease progression. Ammonia was also measured at week 0. For all experiments, rats were under isoflurane anaesthesia (1.5-2% in a mixture of 50%/50% air/$O_2$ for MRS and 1-2% in 100% $O_2$ for PET) with the breathing rate maintained between 60 and 80 resp/minute (SA Instruments, USA). Body temperature was kept between 37.5°C and 38.5°C using a water bath system. All experiments were approved by the Committee on Animal Experimentation for the Canton de Vaud, Switzerland (VD3022.1).

**[1]H MRS:** [1]H MRS experiments were performed on an actively shielded 9.4 Tesla horizontal magnet (Magnex Scientific, UK), 31-cm inner diameter bore, with a Direct Drive console (Varian Inc., USA) and a home-made transmit/receive quadrature surface radio-frequency coil. Anatomical T2-weighted images were acquired in the axial plane to position the volumes of interest (VOIs) for [1]H MRS scans using a multislice turbo-spin-echo sequence (repetition time (TR)/effective echo time ($TE_{eff}$) = 4000/52 ms, echo train length = 8, field of view (FOV) = 23x23 mm$^2$, slice thickness = 1 mm, 15 slices, matrix size =256x256, 1 average). The SPECIAL[26] sequence was used for localized spectroscopy with TE= 2.8 ms, TR= 4 s, 160 averages (10 blocks of 16 averages), a 5 kHz spectral width and 4096 spectral points. [1]H MRS acquisitions were performed on BDL rats before surgery (week 0) and at 6 weeks post-surgery on two brain regions, hippocampus (week 0: n=4, week 6: n=9) and cerebellum (week 0: n=3, week 6: n=4), with a voxel size of 2.8 × 2 × 2 mm$^3$ and 2.5 × 2.5 × 2.5 mm$^3$ (x,y,z on Figure 3), respectively. VAPOR[27] scheme was used for water suppression and FASTMAP[28] for shimming (target water linewidth in the hippocampus: 9-10 Hz, in the cerebellum: 14-17 Hz). Frequency drift and phase corrections between blocks were applied prior to absolute quantification of metabolites with LCModel[29] (version 6.2). An *in vitro* acquired metabolite basis set and the spectrum of macromolecules measured *in vivo*[30,31] were used for LCModel quantification. The water signal from the same voxel was used as internal reference and the metabolite concentrations were derived from the ratio of peak areas, assuming that the water concentration in the voxel was 44.4 M. An exclusion criterion for

individual metabolite concentrations based on relative Cramer Rao Lower Bounds (rejected if CRLB% > 35%) was used. In addition, metabolites were not reported if more than 75% of quantified concentrations over the investigated group were rejected. Since $^1$H MRS acquisitions were performed at week 0, each animal served as its own control for $^1$H MRS results at week 6.

$^{18}$**F-FDG PET:** PET acquisitions on BDL (n=10) and SHAM (n=8) rats at week 6 post-surgery were conducted on a small animal avalanche photodiode detector-based LabPET-4 scanner, with 250-650 keV energy window and 22.2 ns coincidence time window (Advanced Molecular Imaging, Canada). The acquisitions for each individual rat were performed and reconstructed in two steps:

Step (1) - a 45-min dynamic acquisition on the thoracic region of the animal to extract the image-derived arterial input function (AIF) from the vena cava, followed by,

Step (2) - a 15-min static acquisition at labelling steady-state with the rat brain in the FOV of the PET scanner, to further calculate glucose cerebral metabolic rate[21] ($CMR_{glc}$) maps of the brain.

For step (1), a 60 mm-diameter cylindrical field of view (FOV) in the coronal plane and 36.6 mm in the axial direction (31 slices of 1.18 mm thickness) was positioned on the thoracic region of the rat and a bolus of $^{18}$F-FDG (67.6±11.9 MBq) was injected in the tail vein, followed by a saline chase. Coincidence data were acquired in list mode to allow for a flexible reconstruction of time frames. Dynamic radioactivity density maps were quantified in Bq/ml using the LabPET-4 built-in calibration method and reconstructed using the iterative MLEM algorithm (5 iterations), with a time resolution enabling a good characterization of the bolus input function (24 × 5 s, 6 × 30 s, 5 × 120 s, 6 × 300 s)[23]. The inferior vena cava was then identified from the maximum intensity projection (MIP) images during the FDG bolus passage, using the PMOD software environment (PMOD Technologies Ltd.). The AIF was then extracted from the dynamic PET images from step (1) by averaging the activity in Bq/mL of 4 voxels over 4 successive axial slices (total volume: 4.7 mm$^3$) on the vena cava where the flow of $^{18}$F-FDG was observed.

Following step (1), a 10 to 15 min-static acquisition on the brain was performed for step (2) ($0.5 \times 0.5 \times 1.18\ mm^3$ standard voxel size, 60 mm-diameter coronal FOV, 31 axial slices of 1.18 mm thickness) and the quantified images in Bq/mL were reconstructed with a 15-iteration maximum likelihood expectation maximization (MLEM) algorithm[32].

During post processing, the AIF curve from step (1) was first corrected for radioactivity decay and blood versus plasma tracer content[33]. It was then extrapolated from the end of dynamic acquisition up to the central time of the static acquisition used in step (2) based on a bi-exponential fit started at $t = 2\ min$ (i.e. in the decaying phase of the AIF following the chase). In the calculation of the CMR$_{glc}$, trapezoidal integration of the extrapolated AIF curve was used. Brain images were corrected for radioactivity decay from the start of the vena cava acquisition (step (1)).

Finally, 3D maps of CMR$_{glc}$ were reconstructed following the 2-deoxy-D-[$^{14}$C]glucose quantification method of *Sokoloff et al.*[21]. In this method, the CMR$_{glc}$ is obtained from a three-compartment model, represented in Figure 1: a pool 1 of plasma Glc and $^{18}$F-FDG, a pool 2 of intracellular Glc and $^{18}$F-FDG, and a pool 3 of intracellular G6P and $^{18}$F-FDG6P. The following 4 hypotheses were made. First, the static measurement is performed at a sufficiently late time point and in a homogeneous region such that the kinetic rates, the transport rates, the Glc plasma concentration, all intracellular concentrations and the CMR$_{glc}$ rate are constant. In our extension of this method to 3D CMR$_{glc}$ maps from static FDG-PET images, this assumption of homogeneous region applies to the reconstructed voxel. Second, the $^{18}$F-FDG and $^{18}$F-FDG6P concentrations are present in tracer amounts compared to their non-radioactive counterpart. Third, the hydrolysis of G6P to Glc and $^{18}$F-FD6P to $^{18}$F-FDG can be neglected compared to the reverse phosphorylation step. Forth, all brain regions receive a similar amount of tracer and Glc. From these hypotheses, the CMR$_{glc}$ value was derived, describing the rate of G6P utilization in the tissue[21]:

$$CMR_{glc} = \frac{C_i^*(T) - k_1^* e^{-(k_2^*+k_3^*)T} \int_0^T C_p^* e^{+(k_2^*+k_3^*)t} dt}{LC \times \left[\int_0^T \frac{C_p^*(t)}{C_p} dt - e^{-(k_2^*+k_3^*)T} \int_0^T \frac{C_p^*(t)}{C_p} e^{+(k_2^*+k_3^*)t} dt\right]} \quad (1)$$

where $C_i^*$ is the summed concentration of intracellular radioactive compounds ($^{18}$F-FDG6P and $^{18}$F-FDG) i.e. the quantity measured at steady-state in a PET experiment, T the central time of the steady-state static acquisition after the bolus injection, $k_1^*$ the kinetic rate of $^{18}$F-FDG transport from pool 1 to pool 2 through the BBB, $k_2^*$ the reverse $^{18}$F-FDG transport rate from pool 2 to pool 1, $k_3^*$ the $^{18}$F-FDG phosphorylation rate into $^{18}$F-FDG6P, $C_p^*$ the time-dependent plasma $^{18}$F-FDG concentration and $C_p$ the constant plasma Glc concentration. It is assumed that the chemical reaction between FDG and FDG6P is at equilibrium and that partial volume effect (additional radioactivity from the blood vessels measured in the tissue) is negligible when T is large.

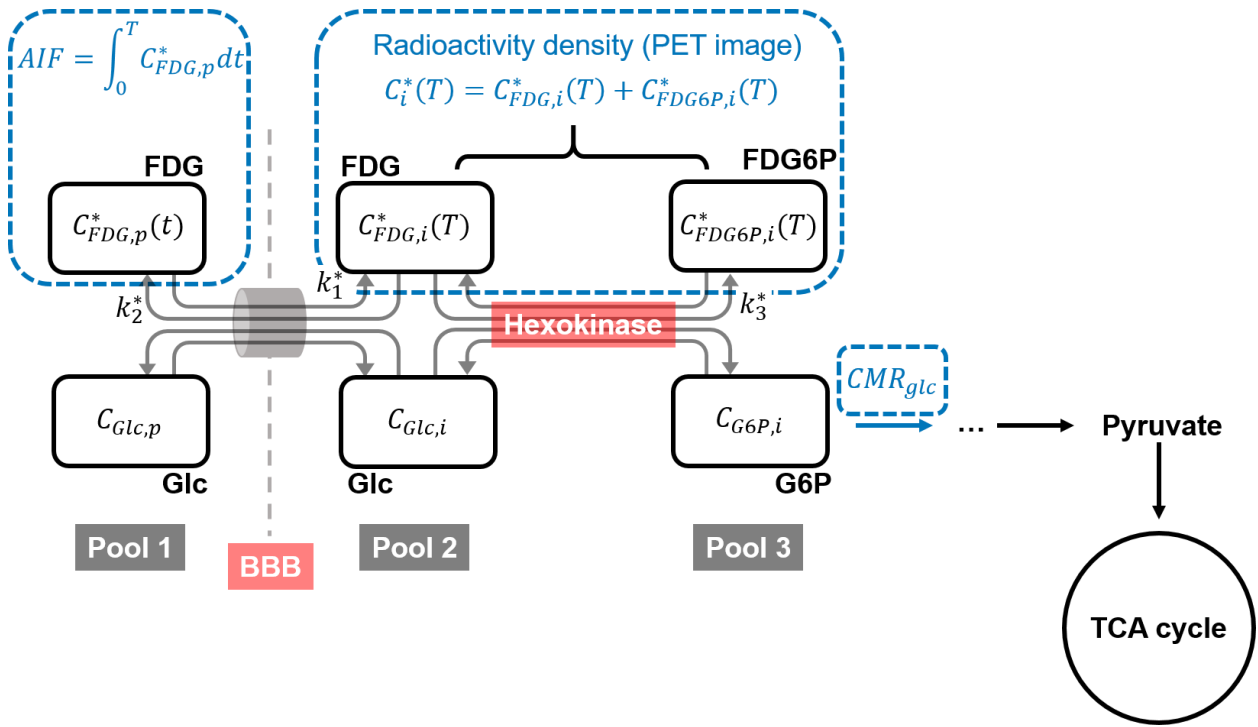

Figure 1 – Three-compartment model of glucose and FDG metabolism used to compute CMR$_{glc}$ values: the kinetic constants, k, and the pool concentrations, C, are denoted with a * when related to the radiolabeled compounds.

If T is large enough (experimentally, 45 min), the equation (1) can be approximated by[21]:

$$CMR_{glc} = \frac{C_i^*(T) \times C_p}{LC \times \int_0^T C_p^*(t)dt} \qquad (2)$$

, where $\int_0^T C_p^*(t)dt$ is the integral of the AIF from step (1) of the *in vivo* acquisition and $C_i^*(T)$ the steady-state brain radioactivity density, as measured from the PET images from step (2). LC is the Lumped Constant, which accounts for the competition between Glc and [18]F-FDG at the transport and phosphorylation steps, as both substrates use the same BBB transporters[34] and are phosphorylated by the hexokinase[35]. In our study, the Lumped Constant (LC) was set to 0.71, as done previously in rat brain studies[36]. Glycemia $C_p$ was measured at the end of step (2) in the tail vein. Since $C_i^*(T)$ is measured for each image voxel from the steady-state acquisition over the brain (step (2)), the derivation of the $CMR_{glc}$ results in a 3D metabolic $CMR_{glc}$ map with same nominal spatial resolution as the PET acquisition itself (i.e. $0.5 \times 0.5 \times 1.18$ mm$^3$), individually for each animal.

**PET-atlas registration:** Since PET images suffer from low spatial resolution and poor contrast, direct PET to atlas registration is challenging. To circumvent this limitation, the MRI anatomical images of one rat were used as an intermediate registration step (see the procedure described in Figure 2). The CMR$_{glc}$ map from one animal with its corresponding MRI anatomical images were chosen as a reference pair. In step A, this reference CMR$_{glc}$ map was registered to its corresponding MRI image using mutual-information-based rigid transformation[37], and in step B, to the Waxholm Space Atlas[38] using affine and nonlinear symmetric deformable registration (SyN) through the Advanced Normalization Tools[39]. In step C, all other individual CMR$_{glc}$ maps were registered to the reference CMR$_{glc}$ map by applying rigid and seven-degrees of freedom similarity transformation with normalized gradient field similarity measure in MeVisLab[40]. Following step C, atlas labels were resampled to each individual PET space to perform a region of interest (ROI)-based analysis, where CMR$_{glc}$ maps were averaged over the hippocampus and the cerebellum regions, respectively.

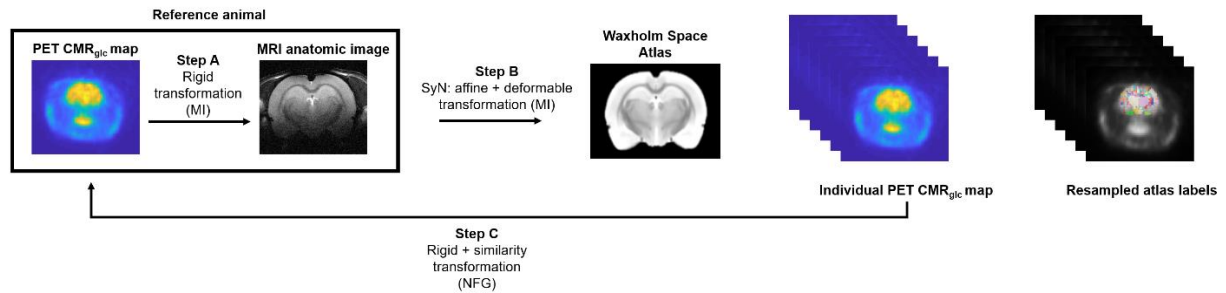

*Figure 2 – PET to atlas registration: the CMR$_{glc}$ map of a reference animal was first registered to its corresponding MRI anatomical images, and then to the Waxholm Space Atlas[38]. Other CMR$_{glc}$ maps were then aligned with the reference one to obtain segmentations for the cerebellum and hippocampus. SyN: nonlinear symmetric deformable registration, MI: mutual information, NFG: normalized field gradient*

**Statistical analysis**: All data are presented as mean ± standard deviations (SD) and assumed to be Gaussian-distributed. Variance equality was tested prior to any statistical test using a Fisher test (F-test of equality of variances). For plasma bilirubin and blood ammonia measurements, an unpaired Student's t-test between week 0 and 6 blood or plasma concentrations was performed. For brain volumes, the mean of the total brain volume covered by all atlas labels derived from the PET to atlas registration was compared between SHAM and BDL rats using an unpaired Student's t-test.

For $^1$H MRS, only Asc concentrations in the hippocampus featured non-equal variances between week 0 and week 6. All metabolites mean concentrations ± SD at week 0 and week 6 were therefore compared using an unpaired Student's t-test, except for Asc concentrations in the hippocampus which was compared with a Welch's t-test, accounting for unequal variances.

For all metabolites where both brain regions displayed significant changes between week 0 and 6 (Gln, Glu and Ins+Tau+tCr+tCho), the impact of the brain region on concentration changes over the weeks was tested through the interaction of the week factor and the brain region factor in a two-way ANOVA (Prism 5.03, Graphpad, La Jolla CA US).

For $^{18}$F-FDG PET, CMR$_{glc}$ variances between SHAM and BDL rats for each brain region were found equal, thus mean CMR$_{glc}$ ± SD were compared using a Student's t-test. We also checked that the standard error on the mean CMR$_{glc}$ over the region of interest was smaller than the SD in the mean CMR$_{glc}$ between animals, retrospectively ensuring that using the SD was meaningful.

The following statistical significance values were used: * $p < 0.05$, ** $p < 0.01$, *** $p < 0.001$, **** $p < 0.0001$.

# Results

**Biochemical measurements**: plasma bilirubin (<0.5 mg/dl at week $0^5$ to 8.07±2.03 mg/dl at week 6, n=1,****) and blood ammonia (89±43 µM at week 0, n=4 to 127±25 µM at week 6, n=4, non-significant) increased in all BDL rats, confirming the induced chronic liver disease (supplementary material, Figure S.1).

**$^1$H MRS - impaired neurometabolic profiles in BDL rats:** Representative spectra acquired in BDL rats at week 0 and 6 in both brain regions are shown in Figure 3. The ultra-short TE allowed the detection of 15 brain metabolites: ascorbate (Asc), glycerophosphocholine (GPC), phosphocholine (PCho), creatine (Cr), phosphocreatine (PCr), GABA, glutamine (Gln), glutamate (Glu), glutathione (GSH), myoinositol (Ins), lactate (Lac), N-acetylaspartate (NAA), Nacetylaspartylglutamate (NAAG), phosphoethanolamine (PE) and taurine (Tau). In the hippocampus, the group analysis showed a strong increase of Gln between week 0 and 6 (+73%,***), and a decrease of Glu (-13%,**), Tau (-20%,**), Ins (-11%, *), total creatine (Creatine+Phosphocreatine, tCr) (-10%,****) and Asc (-17%,**). In the cerebellum, the group analysis showed an increase of Gln, which was even stronger than the one in the hippocampus (+114%,***, with a 1.6-fold significant difference in % change between the two brain regions,*), a decrease of Glu (-22%,**) and Tau (-37%,**), but no significant difference was observed for tCr, Ins and Asc. GABA also showed a significant decrease in the cerebellum (-43%,* p=0.04). Additionally, the main metabolites playing a role in osmoregulation (tCr, tCho, Ins, Tau) were summed to evaluate the osmoregulatory response to the Gln-induced osmotic stress and a significant decrease was observed in the hippocampus (-13%,***) and in the cerebellum (-15%,*), as shown in Figure 5.C. All other individual metabolites that were reliably quantified (GSH, Lac, PE, total *N*-acetylaspartate (tNAA), and total choline (tCho)) showed no significant difference between week 0 and 6 in any of the two brain regions. Alanine (Ala), aspartate (Asp), scyllo-inositol (Scyllo), β-hydroxybutyrate (bHB) and glucose (Glc) were present in the basis set but were excluded from the analysis.

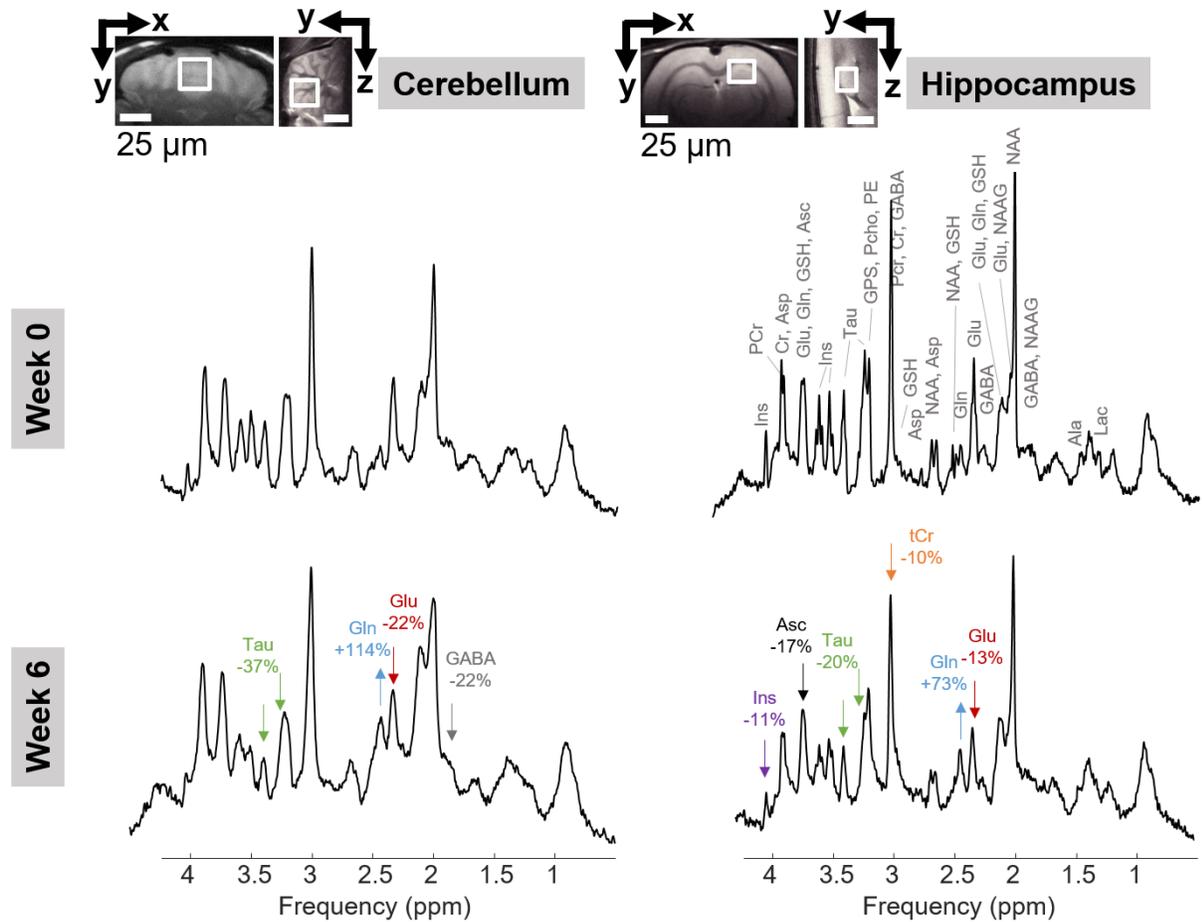

*Figure 3 - <sup>1</sup>H MRS spectra acquired at 9.4 T in the hippocampus and cerebellum of BDL rats at week 0 and 6 post-surgery. Arrows show significant differences observed between week 0 and week 6 for each brain region, and point solely to non-overlapping (or least-overlapping) peaks for the displayed metabolites. The two voxels on anatomical T2-weighted images are shown at the top, where (x,y,z) are the MRI gradient directions.*

[18]**F-FDG - impaired glucose uptake in BDL rats:** The image-derived AIF was reliably measured for each rat from the radiotracer bolus observed in the vena cava during the 45min-dynamic acquisition. Figure 4.A shows a representative AIF prior to correction for blood/plasma FDG content, as well as the chosen VOI over the vena cava based on the maximum intensity projection image. The higher temporal resolution at the beginning of the AIF acquisition allowed for accurate mathematical integration of the input function in the period when the bolus and the chase induced a fast variation of blood FDG activity. The CMR$_{glc}$

derivation from the step (2) PET acquisition over the brain at steady-state enabled the reconstruction of high resolution 3D metabolic maps for individual animals with the same spatial resolution as the reconstructed PET images. A typically two-fold lower $CMR_{glc}$ was observed in BDL versus SHAM rats on all axial slices (Figure 4.B shows an example of $CMR_{glc}$ maps obtained on one BDL and one SHAM rat).

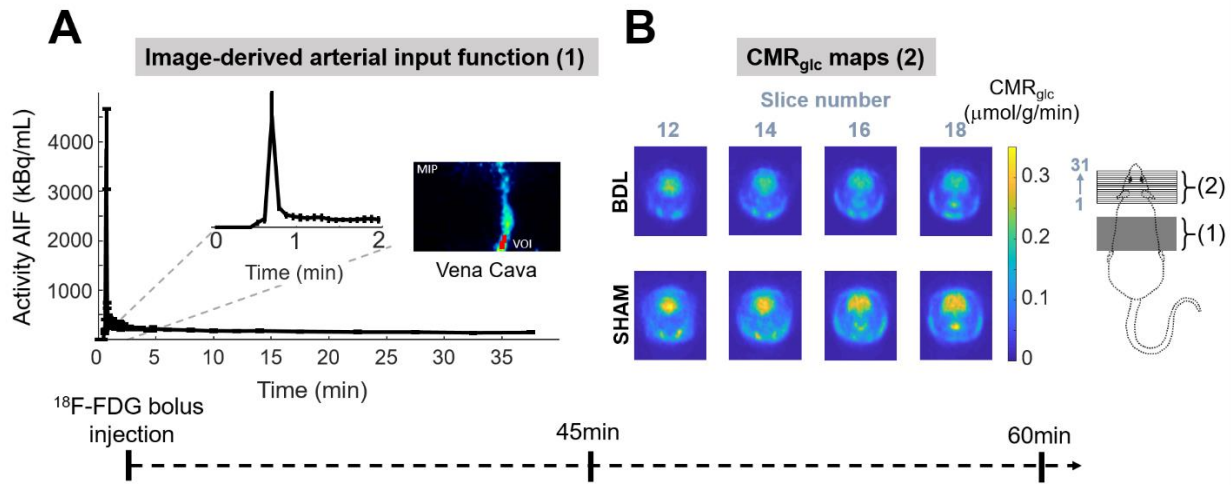

*Figure 4 - $^{18}F$-FDG PET arterial input function and $CMR_{glc}$ maps acquired in BDL and SHAM rats. A. Representative arterial input function curve before correction for blood/plasma content[33], with corresponding VOI in the vena cava. B. Typical $CMR_{glc}$ maps in a BDL and a SHAM rat for four central slices. The rightmost part of the figure shows the position of the PET scanner FOV for the two acquisitions and the slices order. The bottommost part of the figure shows the time repartition of the two acquisitions after injection of the radiotracer.*

PET to atlas registration through MRI anatomic images enabled a ROI-specific measure of glucose uptake and a quantitative comparison between PET and $^1H$ MRS data in the hippocampus and the cerebellum. Figure 5.A shows the atlas labels for the two brain regions. A significant 2.66-fold and 2.53-fold smaller $CMR_{glc}$ in BDL rats compared to SHAM rats (fig. 5.B) was measured respectively in the cerebellum (SHAM: 0.337±0.064 µmol/g/min, BDL: 0.127 ±0.052 µmol/g/min, ****) and in the hippocampus (SHAM: 0.348±0.068 µmol/g/min , BDL: 0.138 ±0.063 µmol/g/min, ****). The proposed co-registration provided quantitative metrics to the differences observed globally in the axial slices of the $CMR_{glc}$ maps (Figure 4.B).

Figure 5.C summarizes the $^1$H MRS results presented in Figure 3, together with the colocalized PET results presented in Figure 4, allowing to draw an overall picture of glucose uptake and neurometabolic profiles alterations happening in BDL rats at week 6 in both brain regions. The BDL rats showed a smaller $CMR_{glc}$ (i.e. smaller glucose uptake) in both regions, an increase in glutamine (cerebellum: +114%, hippocampus: +73%), decrease in Glu (cerebellum: -22%, hippocampus: -13%) and main osmolytes (cerebellum: -15%, hippocampus: -13%), compared to control rats, together with a decrease in some low concentrated metabolites (Asc in the hippocampus (-17%), and in GABA in the cerebellum (-22%). Additionally, Gln increase was significantly stronger in the cerebellum compared to the hippocampus (week 0 to 6 % change), and Glu, Tau and $CMR_{glc}$ show a stronger decrease (although not statistically significant) in the cerebellum.

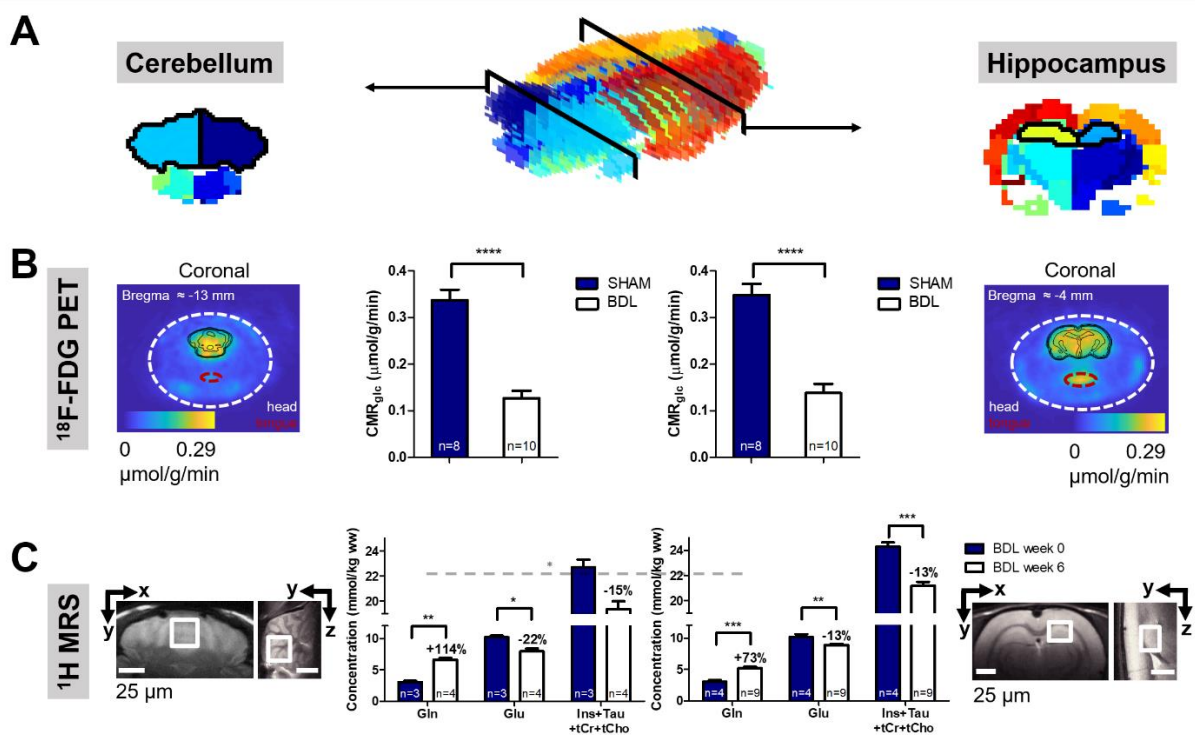

*Figure 5 - Atlas-based co-localization of $^1$H MRS and $^{18}$F-FDG PET. A. Color-coded atlas regions, with the regions of interest (cerebellum and hippocampus) highlighted by solid lines. B. $^{18}$F-FDG PET $CMR_{glc}$ values averaged over the atlas labels of the cerebellum and hippocampus. C. $^1$H MRS metabolites quantifications in a voxel localized in the cerebellum and in the hippocampus. Low concentrated metabolites (Asc, GABA) are not displayed.*

## Discussion

The present study proposes a new quantification method to extract quantitative information on glucose metabolic rate from the raw PET image. This was made possible by using for the first time the combination of the image-derived AIF[23] measurement and the derivation of $CMR_{glc}$ maps using the Sokoloff approach[21], associated with the registration of PET images to a rat brain atlas through MRI T2-weighted anatomic acquisitions. This new approach rendered the 3D PET image fully quantitative and provided an easier experimental implementation with minimal invasiveness of the AIF acquisition as compared with the standard technique employing repeated manual blood samplings. Consequently, we also report here the first *in vivo* study in BDL rats using both $^{18}$F-FDG PET and $^{1}$H-MRS to map brain glucose uptake concomitant with the measurement of brain metabolites alterations, bringing a new insight into brain energy metabolism in HE.

**Consequences of ammonia load on neurometabolic profiles in BDL rats:** The increase in Gln and decrease in osmolytes (Ins, Tau, tCr) observed in the present study are in agreement with a previously published study in the hippocampus of BDL rats[5]. Very interestingly, this important finding was observed here using even fewer animals. Ins, tCr and Tau have been suggested as regulators of cellular volume during induced swelling[41,42], here caused by the Gln load in the astrocytes, and where an efflux could help restoring the osmotic balance.

A primary explanation for decreased Glu in the two brain regions analyzed in this study arises from its excessive use, together with ammonia, for GS-mediated Gln production[5,43]. In addition, altered neurotransmission, both glutamatergic[44] and GABAergic[45,46], has also been proposed as a consequence of chronic hyperammonemia, which could reflect the decrease in Glu observed in both regions and the decrease of GABA in the cerebellum.

The benefit of the spectral separation of Gln/Glu with $^{1}$H MRS spectroscopy at 9.4 T confirms its importance in the context of HE. Indeed, the combined reduction of Glu and increase in Gln concentrations could partially compensate each other when only measuring Glx, the sum of Glu and Gln, a common limitation in $^{1}$H MRS at lower fields. In addition, the cerebellum appears more vulnerable to metabolic changes compared to the hippocampus in BDL rats, confirming previously reported work in this rat model[47].

The Asc decrease in the hippocampus of BDL rats measured in this study is in line with another $^1$H MRS study reported previously in BDL rats[5]. Of note, Asc absolute concentrations should be interpreted with care, as it is a low-concentrated and overlapping metabolite, but its relative decrease between week 0 and 6 is informative. Asc is playing an antioxidant role, therefore its decrease is usually linked to the oxidative stress occurring in the pathogenesis of HE[48] and is also in agreement with one of our previous studies where EPR was used as complementary technique to validate the $^1$H MRS changes[49].

**Impaired energy metabolism in BDL rats:** Our present findings suggest an altered energy metabolism in BDL rats measured with $^{18}$F-FDG PET, in agreement with a Glc hypometabolism observed in a patient with decompensated cirrhosis using FDG PET[18]. In contrast, a previous longitudinal study using $^1$H MRS and $^{31}$P MRS on BDL rats has reported no change in the steady-state concentrations of energy metabolites (i.e. Adenosine triphosphate (ATP), tCr, Lac, while Glc was not reported after week 4) before week 8 post-surgery, with only Adenosine diphosphate (ADP) showing a significant decrease at week 8 post-BDL[8]. This discrepancy can be explained by the different nature of MRS and PET measurements and the complementary information that they provide. While $^1$H MRS measures metabolic pool sizes and reflects the equilibrium changes of biochemical reactions, PET is a kinetic probe that informs on glucose metabolic fluxes. Additionally, it has been shown that brain tissue Glc measured by $^1$H MRS tends to reflect the concentrations of plasma Glc if the later varies sufficiently slowly[50,51], thus informing on Glc homeostasis rather than its metabolism. Glc pools are also challenging to measure using $^1$H MRS because Glc is strongly overlapping with other metabolites on the upfield region of the spectra and with the water residual on the downfield region. Finally, ADP and ATP pools could remain constant if alternative substrate to glucose (such as Lac[10] or ketone bodies[52]) were to be used in the TCA cycle, but more exploratory work in BDL rats is required to test this hypothesis, as well as its link with a potential impairment of the Gln/Glu cycle[9].

**CMR$_{glc}$ versus standardized uptake value (SUV):** Previously published $^{18}$F-FDG PET studies in HE[12–14] in patients and preclinical models show little or no impairment in glucose uptake, whereas we observed a strong difference between BDL and SHAM rats. In addition to the expected difference in HE type (chronic or acute, minimal or overt) and disease characteristics between human and animal studies, we believe that this discrepancy is mainly due to the method used to quantify $^{18}$F-FDG PET data.

While most studies use the SUV (in g/ml, defined as the ratio between the quantified radioactivity density maps of the brain in Bq/ml and the fraction of the injected dose (Bq) over the weight of the animal (g)), the CMR$_{glc}$ is rarely exploited. The SUV is widely used for its robustness and simplicity but is only a semi-quantitative approach, reflecting the normalized density of the tracer distribution in the brain. Its normalization is derived from a macroscopic information (fraction of the injected FDG dose over the weight of the animal) that may overlook subtle changes in the physiology of the animal.

To illustrate this point, the comparison between the CMR$_{glc}$ and the SUV for both groups for the hippocampus and cerebellum is shown in Figure 6.A and B. Both CMR$_{glc}$ and SUV calculation methods start with the same radioactivity density images acquired at FDG labelling steady-state. Figure 6.C shows the normalization terms involved in the CMR$_{glc}$ formula (integral of the AIF and final blood glycemia – $C_p$) and in the SUV formula (injected dose of FDG and weight of the animal). No difference between BDL and SHAM rats in either of the two investigated brain regions can be detected with the SUV (Figure 6.A and B, hippocampus – BDL: 2.03±0.15 g/mL, SHAM: 2.23±0.48 g/mL, cerebellum - BDL: 2.17±0.34 g/mL, SHAM: 1.92±0.53 g/mL). The difference in its macroscopic normalization factors (injected dose and weight) is not statistically significant between BDL and SHAM rats (Figure 6.C, weight - BDL: 328±45 g, SHAM: 363±31 g, dose – BDL: 70.51±10.82 MBq, SHAM: 63.85±12.77 MBq), leading to no difference in the SUV. However, the CMR$_{glc}$ normalization factors (AIF curves and final glycemia) both display a difference between the groups, the latter being significant (BDL: 3.3±1.5 mM, SHAM: 10.1±2.1 mM, ****).

Interestingly, with the same injected dose of FDG in the tail vein for both groups, the average maximum value of the AIF curve from the BDL group is smaller than the one from the SHAM group. This observation suggests that the injected dose is not an accurate measure of the true tracer availability for the brain, since systemic effects, such as the metabolism of other organs, and particularly in this study, of the liver, might affect the blood FDG available for the brain. This physiological effect would have been overlooked using the SUV quantification. The same reasoning holds for the comparison between glycemia in the CMR$_{glc}$ and the weight in the SUV. The latter is also a poor indicator of the physiology of the animal since BDL and SHAM rats have on average the same weight, but BDL rats have a much lower blood glycemia than the SHAM rats, which would have not been taken into account using the SUV.

Because the CMR$_{glc}$ is derived from the kinetics of the 3-compartment model described in Figure 1 and in equations (1) and (2), its expression involves local information on glucose uptake through the ratio between glycemia ($C_p$) and the total amount of tracer in the blood (the summed AIF) multiplied by the LC[21] (see equation (2)). However, the need for a carefully-sampled AIF is often the main difficulty preventing its wider use in metabolic imaging studies. In rodent studies, the small blood volume is a strong limitation for repeated manual sampling. Many technical challenges are also linked to the use of continuous measurements with external blood counters which often result in non-negligible physiological effects on the animal. Additionally, both manual blood sampling and external counters require the cannulation of veins or arteries, often rendering the experiment terminal. On the contrary, the proposed approach with the image-derived AIF provides a practical alternative to the manual blood sampling or external blood counters. It renders similar results[23] with particularly high temporal resolution, makes the measurement less invasive and allows for longitudinal studies with the same animal.

To further ensure a fair comparison between the groups, brain volumes were compared between BDL and SHAM rats and are presented in the supplementary material (Figure S.2). No overall brain atrophy was observed in BDL rats compared to SHAM rats (BDL brain volume – 1959.4±41.0 mm$^3$, SHAM brain volume – 1939.2±48.6 mm$^3$), ensuring that a given amount of tracer/plasma Glc is used by the same amount of brain tissue between the two groups.

Finally, the LC in the CMR$_{glc}$ formula, in conjunction with the glycemia, also mirrors an important physiological aspect as it accounts for the competition between glucose and FDG and their respective affinity for blood brain barrier transporters and hexokinase. In particular in the present study, the glycemia values strongly differ between the BDL and SHAM groups. All kinetic constants from the 3-compartment model, both for enzyme-mediated transport through the BBB and the enzyme-mediated phosphorylation for both substrates are described by a Michaelis-Menten equation modified to account for competitive substrates and mutual inhibition[53]. This competition is formally contained in the Lumped Constant expression from *Sokoloff et al*[21]:

$$LC = \frac{1}{\Phi} \times \frac{k_1^*/(k_2^* + k_3^*)}{k_1/(k_2 + k_3)} \times \frac{V_m^*/K_m^*}{V_m/K_m} \quad (3)$$

, where $\Phi$ is the fraction of G6P that will be further metabolized in the glycolysis, the second fraction describing the ratio of kinetic constants of the radiotracer over the ones of natural Glc, and the last part of the fraction the ratio between Michaelis-Menten constants, $V_m^{(*)}$ the maximum velocity and $K_m^{(*)}$ Michaelis-Menten constants for the hexokinase reaction of either Glc or FDG. The LC constants were assumed identical for both groups. Importantly, the difference in blood glycemia and the resulting differential distribution of blood glucose versus FDG between groups would have been overlooked when analyzing the glucose uptake with the SUV approach, for which we observed no significant difference in animal weights between the two groups.

For all the reasons mentioned above, when practically feasible, we suggest using a quantitative approach for the analysis of FDG uptake which enables the determination of the CMR$_{glc}$ in studies involving group comparison where the physiology of the animal could greatly vary.

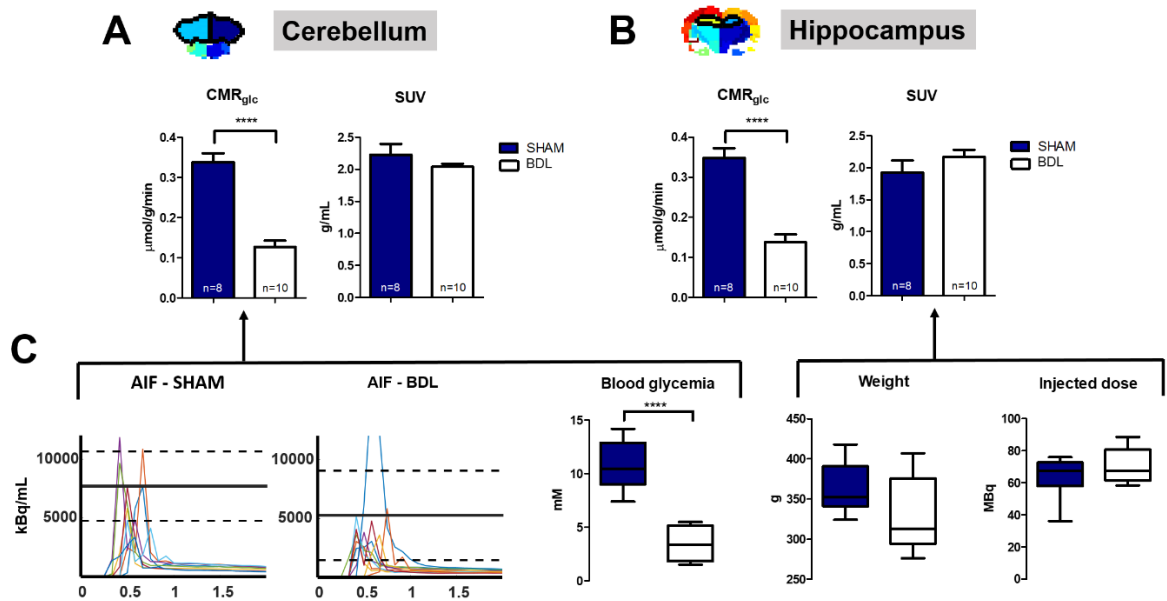

*Figure 6 - Comparison between two different metrics for PET data, the fully-quantitative $CMR_{glc}$ and the semi-quantitative SUV for the cerebellum (A) and the hippocampus (B) (color-coded atlas regions, with the regions of interest (cerebellum and hippocampus) highlighted by solid lines) and its respective normalization terms (C). Individual arterial input function curves from each animal are displayed in the left part of panel C, the full line being the mean of the maximum of the AIF and the dashed line the corresponding mean±SD.*

## Conclusions

We report the first multimodal [18]F-FDG PET and localized [1]H MRS *in vivo* study in the type C HE BDL rat model. In addition, a refined FDG-PET analysis was developed to determine a quantitative and regional measurement of the cerebral metabolic rate of glucose in terms of 3D $CMR_{glc}$ maps, based on an image-derived AIF. It revealed a 2-fold lower glucose uptake in the hippocampus and cerebellum of BDL versus SHAM rats. A concomitant increase in glutamine, a decrease in glutamate and in the osmolytes was measured with localized [1]H MRS in the same brain regions. This novel finding reopens the debate of energy failure in the pathophysiology of type C HE.

## Acknowledgements

Supported by the SNSF project No 310030 173222 and the European Union's Horizon 2020 research and innovation program under the Marie Sklodowska-Curie grant agreement No 813120 (INSPiRE-MED). The authors would like to thank Dr. Corina Berset (CIBM) for her work in pilot PET experiments and the veterinary team at CIBM for support during experiments. This work was made possible thanks to the CIBM Center for Biomedical Imaging, a Swiss research center of excellence founded and supported by Lausanne University Hospital (CHUV), University of Lausanne (UNIL), Ecole Polytechnique Fédérale de Lausanne (EPFL), University of Geneva (UNIGE) and Geneva University Hospitals (HUG).

## Supplementary material

**Increased bilirubin and ammonia: markers of the induced HE**

Figure S.1 shows the expected increase in plasma bilirubin (<0.5 mg/dl at week $0^5$ to 8.07±2.03 mg/dl at week 6, n=1,****) and blood ammonia (89±43 µM at week 0, n=4 to 127±25 µM at week 6, n=4, non-significant) observed in BDL rats over weeks post-surgery, confirming the type C HE induced after the bile duct ligation surgery.

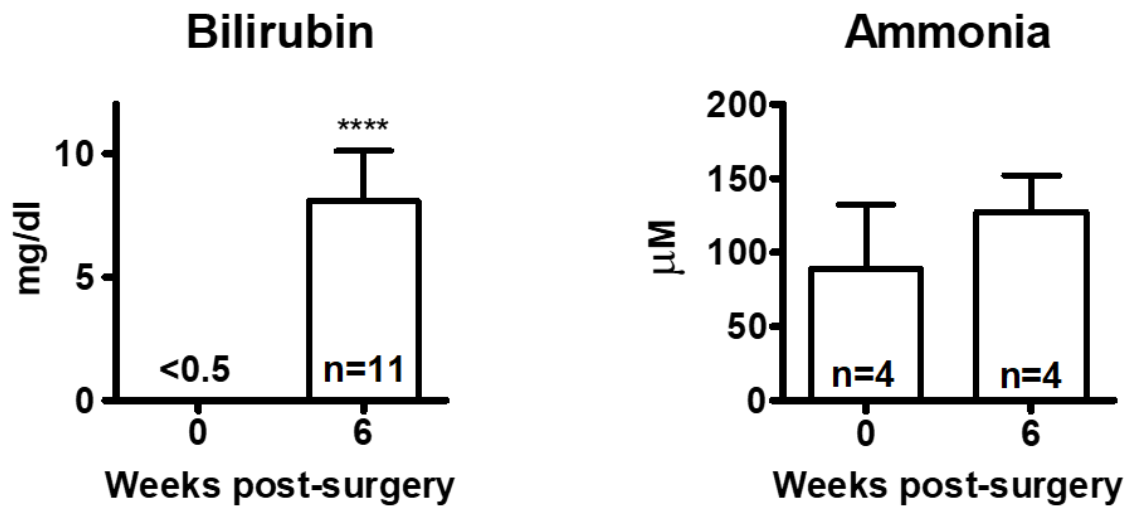

*Figure S.1 – Evolution of plasma bilirubin and blood ammonia in BDL rats over weeks post-surgery*

**No brain atrophy in BDL rats**

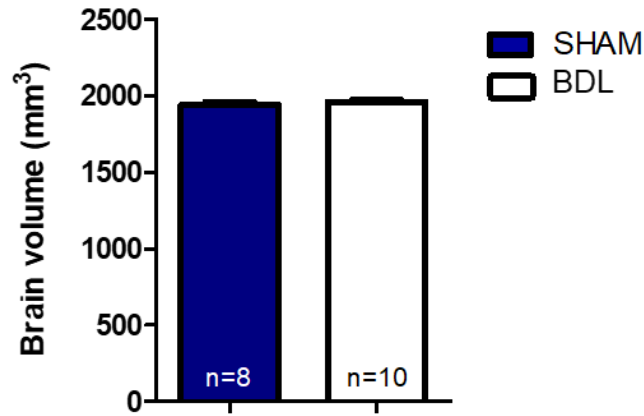

*Figure S.2 – Brain volume comparison between BDL and SHAM rats*

Atlas-based PET registration allows to compute the brain volumes by summing the voxels from all brain labels. Figure S.2 shows that no brain atrophy is observed in BDL rats compared to SHAM rats (BDL brain volume – 1959.4±41.0 mm$^3$, SHAM brain volume – 1939.2±48.6 mm$^3$), thus confirming that no bias is introduced in the comparison of CMR$_{glc}$ values between the two groups.